\documentclass{PoS}

\usepackage{mathtools,siunitx,booktabs,braket,slashed}
\usepackage{caption}
\usepackage{tikz}

\title{Position-space approach to hadronic light-by-light scattering in the muon 
$g-2$ on the lattice}

\ShortTitle{Position-space approach to HLbL scattering in the muon $g-2$ on the lattice}

\author{\speaker{Nils Asmussen},
   Jeremy Green,
   Harvey B.\ Meyer
   and Andreas Nyffeler\\
{PRISMA Cluster of Excellence,  Institut f\"ur Kernphysik and Helmholtz Institute Mainz, 
 Johannes~Gutenberg-Universit\"at~Mainz, 55099 Mainz, Germany}\\
  E-mail: \email{\{asmussen,
    green,
    meyerh,
    nyffeler\}@kph.uni-mainz.de}
}

\abstract{The anomalous magnetic moment of the muon currently exhibits a 
discrepancy of about three standard deviations between the experimental value and recent Standard Model 
predictions. The theoretical uncertainty is dominated by the hadronic vacuum 
polarization and the hadronic light-by-light (HLbL) scattering contributions, 
where the latter has so far only been fully evaluated using different models. To 
pave the way for a lattice calculation of HLbL, we present an expression for the 
HLbL contribution to $g-2$ that involves a multidimensional integral over a 
position-space QED kernel function in the continuum and a lattice QCD four-point 
correlator. We describe our semi-analytic calculation of the kernel and test the 
approach by evaluating the $\pi^0$-pole contribution in the continuum.}

\FullConference{34th annual International Symposium on Lattice Field Theory\\
                 24-30 July 2016\\
                 University of Southampton, UK}

\DeclarePairedDelimiter\abs{\lvert}{\rvert}
\DeclareMathOperator\trace{Tr}

\newcommand\hlbl{HLbL}
\newcommand\dd{\mathrm{d}}

\newcommand\mcI{\mathcal{I}}
\newcommand\mcL{\mathcal{L}}
\newcommand\mcG{\mathcal{G}}
\newcommand\bmcL{\bar{\mathcal{L}}}
\newcommand\epsavg[1]{\langle{#1}\rangle_{\hat\epsilon}}
\newcommand\order[1]{\mathcal{O}(#1)}

\newcommand\etal{\emph{et al.}}

\date{July 26, 2016}

\begin{document}

\section{Introduction}
One of the most precisely measured physical quantities is the anomalous magnetic 
moment of the muon \(a_\mu=\frac{g_\mu-2}{2}\). The Standard Model provides 
predictions of similar accuracy as the measurements. Comparing the theoretical 
and experimental value leads to a very stringent test of the Standard Model. 
A long-standing discrepancy of three 
standard deviations or more is observed in \(a_{\mu}\), 
\begin{align}
a_\mu=\begin{cases}
\num{116592089+-63e-11}&{\rm experiment~}\cite{PDG2014,Bennett:2006fi}\\
\num{116591790+-65e-11}&{\rm theory~}\cite{Jegerlehner:2009ry}.
\end{cases}
\end{align}

To reduce the uncertainty, experiments planned at Fermilab 
 and at J-PARC~\cite{Hertzog:2015jru} aim to improve the experimental uncertainty by a 
factor of four. To profit most from these efforts, the theoretical uncertainty 
must be reduced by a similar amount.
Despite not being the largest contributions to \(a_{\mu}\), the hadronic vacuum 
polarization (HVP) (\(\order{\alpha^2}\)) and hadronic light-by-light scattering 
contribution (\hlbl) (\(\order{\alpha^3}\)) contribute most to the theoretical 
uncertainty of \(a_{\mu}\).



Unlike the HVP, the \hlbl{} is not fully related to any cross section. The 
estimates of the \hlbl{} rely on models, which leads to large uncertainties. In 
the phenomenological treatment one tries to reduce model uncertainties for the 
dominant contributions (\(\pi^0\,,\,\eta\,,\,\eta^\prime\,;\,\pi\pi\)) by using 
experimental input with the help of dispersion relations, see Colangelo \etal{} 
\cite{Colangelo:2014dfa,Colangelo:2014pva,Colangelo:2015ama} and 
Pauk and Vanderhaeghen~\cite{Pauk:2014rfa}.  Lattice QCD can provide tests of 
dispersive approaches~\cite{Green:2015sra} and also 
a direct first-principle estimate of $a_\mu^{\rm HLbL}$~\cite{Blum:2014oka}.


\section{An expression for $a_\mu^{\rm HLbL}$ adapted to lattice QCD calculations}

\begin{figure}
\centering
\includegraphics{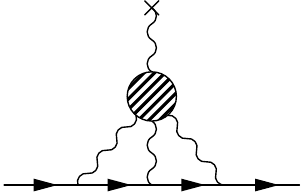}
\caption{Hadronic light-by-light scattering diagram}
\label{fig:muonhlbl}
\end{figure}

The basic idea of our method is to treat the four-point function represented by the
blob in Fig.~\ref{fig:muonhlbl} in lattice regularization, while for
the remaining (QED) parts, we use continuum, Euclidean position-space
perturbation theory in infinite
volume~\cite{DPG2015,Green:2015mva}. In this way, we avoid having
power-law corrections in the volume on $a_\mu^{\rm HLbL}$.

In Euclidean space, the matrix element of the electromagnetic current 
at momentum transfer $k=p'-p$ can be decomposed into form factors as
\begin{align}
\begin{split}
   \bra{\mu^-(p',s')}j_\rho(0)\ket{\mu^-(p,s)}
   &=-\bar{u}^{s'}(p')
\Big[  \gamma_\rho F_1(k^2)+\frac{\sigma_{\rho\tau}k_\tau}{2m}F_2(k^2) \Big]
   u^s(p),
\end{split}
\end{align}
The anomalous magnetic moment is defined as the Pauli form factor at \(k^2=0\), 
\(a_{\mu}=F_2(0)\).
An explicit projection of the vertex to \(a_\mu\) can be given~\cite{Aldins:1970id},
\begin{align}
   F_2(0)=\frac{-i}{48m}\trace\{[\gamma_\rho,\gamma_\tau](-i\slashed{p}+m)\Gamma_{\rho\tau}(p,p)(-i\slashed{p}+m)\}
   \label{eq:KinoshitaTrace}
   \,,
\end{align}
where the HLbL contribution to the vertex reads ($\int_{q}\equiv \int \frac{d^4q}{(2\pi)^4}$, $\int_x \equiv \int d^4x$)
\begin{align}
\begin{split}
& \Gamma_{\rho\sigma}(p',p)=
   -e^6\int_{q_1,q_2} 
   \frac{1}{q_1^2q_2^2(q_1+q_2-k)^2}
   \frac{1}{(p'-q_1)^2+m^2}\frac{1}{(p'-q_1-q_2)^2+m^2}
   \\& \qquad \quad 
   \Big(\gamma_\mu(i\slashed{p}'-i\slashed{q}_1-m)
   \gamma_\nu(i\slashed{p}'-i\slashed{q}_1-i\slashed{q}_2-m)
   \gamma_\lambda\Big)
   \frac{\partial}{\partial k_\rho}\Pi_{\mu\nu\lambda\sigma}(q_1,q_2,k-q_1-q_2),
\end{split}
\\
& \Pi_{\mu\nu\lambda\sigma}(q_1,q_2,q_3)=
\int_{x_1,x_2,x_3}e^{-i(q_1x_1+q_2x_2+q_3x_3)}\Big\langle j_\mu(x_1)j_\nu(x_2)j_\lambda(x_3)j_\sigma(0)\Big\rangle.
\end{align}
We now parametrize the on-shell muon momentum by a unit-vector $\hat\epsilon$ as
\begin{align}
   p=im\hat\epsilon\quad (p^2=-m^2)
   \,,
\end{align}
and write the expression in terms of position-space functions,
\begin{align}
\Gamma_{\rho\sigma}(p,p)\;=&\; -e^6\int_{x,y}
\widehat\Pi_{\rho;\mu\nu\lambda\sigma}(x,y)
K_{\mu\nu\lambda}(x,y,p)\,,
\label{eq:gammarhosigmakernel}
\\
K_{\mu\nu\lambda}(x,y,p)
\;=&\;\gamma_\mu(i\slashed{p}+\slashed{\partial}^{(x)}-m)
\gamma_\nu(i\slashed{p}+\slashed{\partial}^{(x)}+\slashed{\partial}^{(y)}-m)
\gamma_\lambda\mathcal{I}(\hat\epsilon,x,y)\,,
\\
\mathcal{I}(\hat\epsilon,x,y)
\;=&\;\int_{q,k}\frac{1}{q^2\,k^2\,(q+k)^2}\,\frac{1}{(p-q)^2+m^2}\,\frac{1}{(p-q-k)^2+m^2}\;
e^{-i(qx+ky)}\,,
\\
\widehat\Pi_{\rho;\mu\nu\lambda\sigma}(x,y)\;=&\;\int_{z}
iz_\rho\; \Big\langle{j_\mu(x)j_\nu(y)j_\sigma(z)j_\lambda(0)}\Big\rangle
\,.
\end{align}
Since the scalar function \(\mcI\) is logarithmically infrared divergent, we 
need to introduce a regulator. However the derivatives in \(K_{\mu\nu\lambda}\) 
remove the divergence, so that the latter tensor is infrared finite.

Our task is now to evaluate the Fourier integral in \(\mcI\). We express it in 
terms of position-space scalar propagators \(G_m(x)\),
\begin{align}
\mcI(\hat\epsilon,x,y)&=\int_{u,\text{IR-reg}} 
G_0(u-y)J(\hat\epsilon,u)J(\hat\epsilon,x-u)
\,,\\
J(\hat\epsilon,y) &=\int_xG_0(x+y)e^{-m\hat\epsilon\cdot x}G_m(x)
\,,\\
G_m(x)&=\frac{m}{4\pi^2\abs x}K_1(m\abs x)\quad
(K_1\text{ is a modified Bessel function})
\,.
\end{align}
We insert Eq.~\eqref{eq:gammarhosigmakernel} into~\eqref{eq:KinoshitaTrace},  
evaluate the trace and obtain an expression of the form
\begin{align}
a_\mu^\text{\hlbl}=\hat F_2(0)&=\frac{me^6}{3}\int_y
\int_x\mcL_{[\rho,\sigma];\mu\nu\lambda}(\hat\epsilon,x,y)\;i\,\widehat\Pi_{\rho;\mu\nu\lambda\sigma}(x,y)
\,.
\end{align}
We exploit the invariance of $\hat F_2(0)$ under O(4) rotations of the muon momentum 
and average the kernel over the direction of \(\hat\epsilon\),
\begin{align}
\bmcL_{[\rho,\sigma];\mu\nu\lambda}(x,y)=
\frac{1}{2\pi^2}\int\dd\Omega_\epsilon\;
\mcL_{[\rho,\sigma];\mu\nu\lambda}(\hat\epsilon,x,y)
\equiv \Big\langle \mcL_{[\rho,\sigma];\mu\nu\lambda}(\hat\epsilon,x,y)\Big\rangle_{\hat\epsilon}
\,.
\end{align}
Thus we  arrive at the master formula~\cite{Green:2015mva},
\begin{align}
a_\mu^\text{\hlbl}=F_2(0)&=\frac{me^6}{3}\int_y
\int_x\bmcL_{[\rho,\sigma];\mu\nu\lambda}(x,y)i\,\widehat\Pi_{\rho;\mu\nu\lambda\sigma}(x,y)
\,.
\label{eq:master}
\end{align}
The angular average is performed analytically 
by first expanding the dependence of \(J(\hat\epsilon,y)\) on $\hat\epsilon\cdot\hat y$ 
 in Chebyshev polynomials of the second kind $U_n$ ---  a special case of the Gegenbauer expansion,
\begin{align}
\begin{split}
J(\hat\epsilon,y)&=\sum_{n\ge0}z_n(y^2)\;U_n(\hat\epsilon\cdot\hat y).
\end{split}
\end{align}
The coefficients $z_n$ turn out to be linear combinations of products of two modified Bessel functions.
The orthogonality property $\langle U_n(\hat\epsilon\cdot\hat x)\; U_m(\hat\epsilon\cdot\hat y) \rangle_{\hat\epsilon}
=\frac{\delta_{nm}}{n+1}U_n(\hat x\cdot \hat y)$ can then be exploited.

The formula in Eq. (\ref{eq:master}) shows the QED kernel function 
\(\bmcL_{[\rho,\sigma];\mu\nu\lambda}\), that weights the position-space QCD 
four-point function \(\widehat\Pi_{\rho;\mu\nu\lambda\sigma}(x,y)\). 
The kernel has been averaged over the direction of the muon momentum, so that after contracting the 
Lorentz indices the integration reduces to a 3-dimensional integration over 
\(x^2, y^2\) and the angle $\beta$ between $x$ and $y$ (i.e.\ $x\cdot y = |x||y|\cos\beta$),
\begin{align}
   \int_y\rightarrow 2\pi^2\int_0^\infty\dd\abs y\abs y^3\,,\qquad
   \int_x\rightarrow 4\pi\int_0^\infty\dd\abs
   x\,\abs x^3\int_0^\pi\dd\beta
   \sin^2\beta\,.
\end{align}
In the lattice implementation, we plan to make use of the reduction in the
integration dimensionality only for the $y$ integral (see section \ref{sec:lat}).
In analytic calculations, the full reduction ought to be exploited. It is also worth noting that the contraction of 
$\widehat\Pi_{\rho;\mu\nu\lambda\sigma}(x,y)$ with the tensors $\mcG^A_{\delta\rho\sigma\mu\alpha\nu\beta\lambda}$
introduced in Eq.\ (\ref{eq:kernelT}) projects the QCD four-point function onto a rank-three tensor,
which can be decomposed into a smaller number of tensor structures.
Finally, the Bose symmetry of the internal vertices,
\begin{equation}
\widehat\Pi_{\rho;\mu\nu\lambda\sigma}(x,y) = \widehat\Pi_{\rho;\nu\mu\lambda\sigma}(y,x) = 
\widehat\Pi_{\rho;\lambda\nu\mu\sigma}(-x,y-x)\,,
\end{equation}
could be used to further symmetrize the kernel $\bmcL_{[\rho,\sigma];\mu\nu\lambda}(x,y)$.

\section{Semi-analytic calculation of the kernel}
We decompose the QED kernel function 
\(\bmcL_{[\rho,\sigma];\mu\nu\lambda}(x,y)\) into tensors 
\(T_{\alpha\beta\delta}^A(x,y)\):
\begin{align}
\bmcL_{[\rho,\sigma];\mu\nu\lambda}(x,y)
=\sum_{A=I,II,III}\mcG^A_{\delta\rho\sigma\mu\alpha\nu\beta\lambda}T_{\alpha\beta\delta}^A(x,y)
\,.
\label{eq:kernelT}
\end{align}
The \(\mcG^{A}_{\delta\rho\sigma\mu\alpha\nu\beta\lambda}\) are sums of 
products of Kronecker deltas coming from traces of Dirac matrices.
The tensors in turn are decomposed into a scalar \(S\), a vector \(V_\delta\) 
and a tensor part \(T_{\beta\delta}\),
\begin{align}
T_{\alpha\beta\delta}^I(x,y)&=
\partial_\alpha^{(x)}(\partial_\beta^{(x)}+\partial_\beta^{(y)})V_\delta(x,y)\,,\notag\\
T_{\alpha\beta\delta}^{II}(x,y)&=
m\partial_\alpha^{(x)}(T_{\beta\delta}(x,y)+\frac{1}{4}\delta_{\beta\delta}S(x,y))\,,\label{eq:tensorsT}\\
T_{\alpha\beta\delta}^{III}(x,y)&=
m(\partial_\beta^{(x)}+\partial_\beta^{(y)})(T_{\alpha\delta}(x,y)+\frac{1}{4}\delta_{\alpha\delta}S(x,y))\notag
\,.
\end{align}
These parts are given in terms of the function $\mcI$ by
\begin{align}
&\text{scalar:}&S(x,y)=&\epsavg{\mcI}\quad\text{(IR regulated)}\,,\notag\\
&\text{vector:}&V_\delta(x,y)=&\epsavg{\hat\epsilon_\delta\mcI}\,,\\
&\text{tensor:}&T_{\beta\delta}(x,y)=&\epsavg{(\hat\epsilon_\delta\hat\epsilon_\beta-\frac{1}{4}\delta_{\delta\beta})\mcI}\,.\notag
\end{align}
This procedure allows us to parametrize the QED kernel by six form factors, each one a function of \(x^2,y^2,x\cdot y\),
\begin{align}
\begin{split}
S(x,y)=&g^{(0)}\,,
\qquad
V_\delta(x,y)=x_\delta\, g^{(1)}+y_\delta\, {g^{(2)}}\,,\\
T_{\alpha\beta}(x,y)=&
(x_\alpha x_\beta-\frac{x^2}{4}\delta_{\alpha\beta})\,l^{(1)}
+(y_\alpha y_\beta-\frac{y^2}{4}\delta_{\alpha\beta})\,l^{(2)}
+(x_\alpha y_\beta+y_\alpha x_\beta-\frac{x\cdot 
y}{2}\delta_{\alpha\beta})\,l^{(3)}\,.
\end{split}
\label{eq:g012l123}
\end{align}

We have computed all six form factors and stored their values on a 
(\(|x|,\,|y|,\,\cos\beta\)) grid. In that way, the kernel
\(\bar{\mcL}_{[\rho,\sigma];\mu\nu\lambda}\) can be obtained from the form
factors by a cheap computation based on Eqs.~\eqref{eq:kernelT},
\eqref{eq:tensorsT} and \eqref{eq:g012l123}.  We achieved a precision
on the form factors of about five digits using Gaussian quadrature
methods. For instance, the vector form factor \(g^{(2)}\) is given by
\begin{align}
\MoveEqLeft[2]
g^{(2)}(x^2,x\cdot y,y^2)
=
\frac{1}{8\pi y^2\abs{x}\sin^3\beta}
{\int_0^{\infty} du}\, u^2
{\int_0^\pi d\phi_1}
\left\{2\sin\beta+\left(\frac{y^2+u^2}{2\abs{u}\abs{y}}-\cos\beta\cos\phi_1\right)\frac{\log\chi}{\sin\phi_1}\right\}
\notag\\
&\sum_{n=0}^{\infty}
\big\{
	z_n(\abs{u})z_{n+1}(\abs{x-u})
	\left[
		\abs{x-u}\cos\phi_1\frac{U_n}{n+1}
		+
		(\abs{u}\cos\phi_1-\abs{x})\frac{U_{n+1}}{n+2}
	\right]
\notag\\
&	+
	z_{n+1}(\abs{u})z_n(\abs{x-u})
	\left[
		(\abs{u}\cos\phi_1-\abs{x})\frac{U_n}{n+1}
		+
		\abs{x-u}\cos\phi_1\frac{U_{n+1}}{n+2}
	\right]
\big\},
\end{align}
where $\;\abs{x-u}=(\abs{x}^2+\abs{u}^2-2\abs{x}\abs{u}\cos\phi_1)^{1/2}\;$ and 
\begin{align}
\begin{split}
\chi=&\frac{y^2+u^2-2\abs{u}\abs{y}\cos(\beta-\phi_1)}{y^2+u^2-2\abs{u}\abs{y}\cos(\beta+\phi_1)}
\,,\qquad U_n=U_n\Big(\frac{\abs{x}\cos\phi_1-\abs{u}}{\abs{u-x}}\Big)\,.
\end{split}
\end{align}
As can be seen in the left panel of Fig.~\ref{fig:amuNumericalTest}, the form factor $g^{(2)}$ is a smooth function
of its arguments.

\section{Numerical test of the kernel and aspects of the lattice implementation\label{sec:lat}}
\begin{figure}
\centerline{\includegraphics[width=0.485\textwidth]{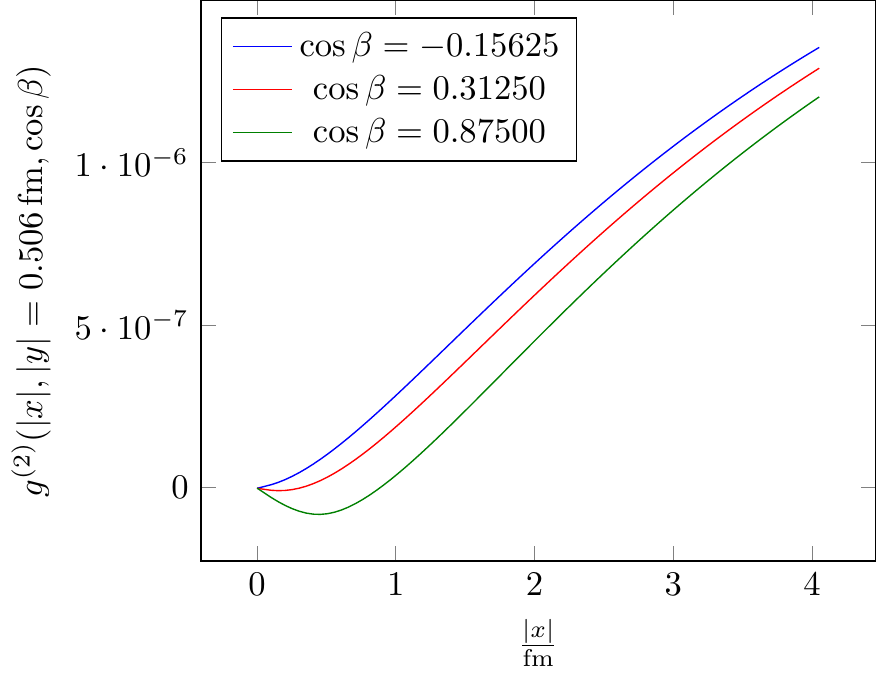}
\includegraphics[width=0.51\textwidth]{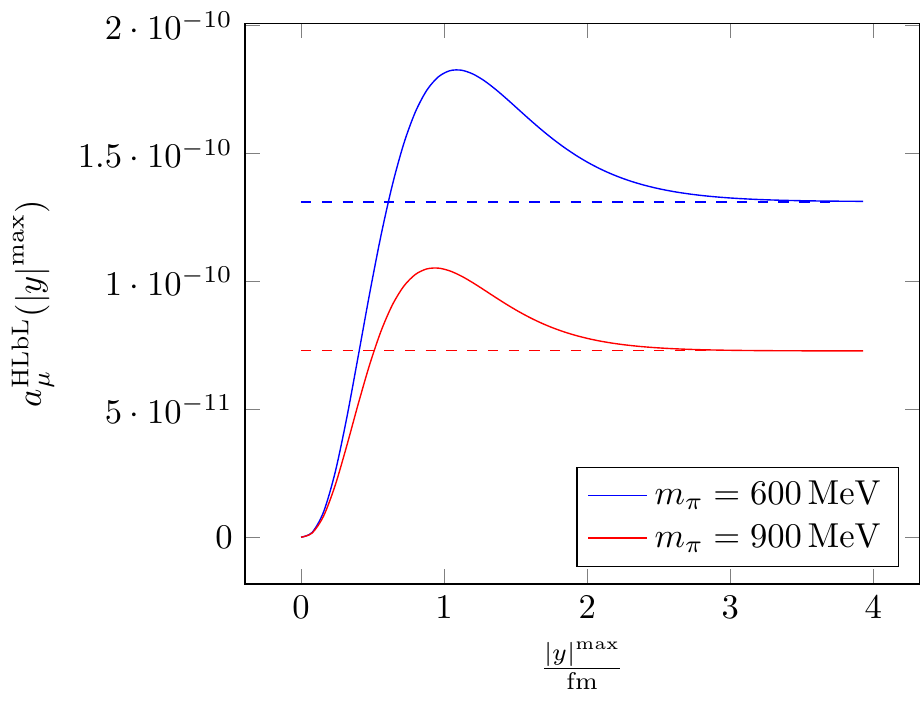}}
\caption{Left: Form factor $g^{(2)}$ for fixed $|y|$ and three different values of 
  $\cos\beta$. Right: Numerical test of the kernel $\bar{\mcL}_{[\rho,\sigma];\mu\nu\lambda}(x,y)$,
computing the pion-pole contribution to $a_\mu^{\rm HLbL}$  for two different  
 pion masses, with an upper bound $\abs{y}^{\text{max}}$ for the integral over 
 $\abs{y}$.  The dashed lines represent the value computed with momentum-space 
 methods~\cite{Knecht:2001qf}.}
\label{fig:amuNumericalTest}
\end{figure}

In order to verify the correctness of the QED kernel, we calculated the pion-pole contribution to
the four-point function \(\widehat\Pi_{\rho;\mu\nu\lambda\sigma}(x,y)\) using  the VMD 
model (as defined in~\cite{Knecht:2001qf}) for the pion transition form factor.  We then computed 
\(a_\mu^{\text{\hlbl},\pi^0}\) using Eq.~(\ref{eq:master}) and compared with the 
result obtained via momentum-space integration~\cite{Knecht:2001qf}.  
Fig.~\ref{fig:amuNumericalTest} (right panel) shows a comparison for two different pion masses. 
Although the integrals over $|x|$ and $|y|$ are expected to be exponentially convergent at large distances, 
numerically the convergence is only achieved for \(\abs y^{\text{max}}\geq 2-\SI{3}{fm}\) even though the pion mass 
\(m_\pi=600-\SI{900}{MeV}\) is quite large. The cutoff for the \(x\) integration 
was fixed at \(\abs x^{\text{max}}=\SI{4.05}{fm}\).

In a lattice QCD implementation, the cost of computing the diagrams
with fully connected quark lines can be estimated as follows.  For
fixed $y$, the \(\dd^4x\) integral is evaluated as the sum $a^4\sum_x$
over all lattice sites with the help of sequential propagators.  If
the remaining one-dimensional integral over \(\abs y\) is done with
$N$ evaluations of the integrand, we need $(1+N)$ forward propagators
and $6(1+N)$ sequential propagators.  We expect to require $N\approx
20$ for a reliable evaluation of the $|y|$ integral.

\section{Summary and outlook}
We presented an explicit formula for \(a_\mu^{\text{\hlbl}}\), Eq.\ (\ref{eq:master}),
consisting of a QED kernel function weighting the Euclidean
position-space QCD four-point function~\cite{DPG2015,Green:2015mva}. We have completed the calculation of  the QED kernel in
the continuum in infinite volume. This computational strategy avoids power-law
finite-size effects.  We exploited the invariance under rotations of
the direction of the muon momentum \(\hat\epsilon\), by averaging
$a_\mu^{\rm HLbL}$ over \(\hat\epsilon\). As a result, only the
one-dimensional \(\abs{y}\) integral must be sampled
stochastically. We computed the form factors parametrizing the kernel
and stored them on disk, so that the kernel can be computed in a
negligible amount of time during the lattice simulation. The
correctness of the position-space approach has been tested by
computing the pion-pole contribution to $\widehat
\Pi_{\rho;\mu\nu\lambda\sigma}(x,y)$ with a VMD form factor and reproducing known results on
$a_\mu^{{\rm HLbL},\pi^0}$ by performing the integrals over $x$ and
$y$.  This type of calculation may also be used in the future to correct for
the leading finite-size effect on $a_\mu^{{\rm HLbL}}$ in our approach, if the pion
transition form factor is computed along the lines of~\cite{Gerardin:2016cqj}.

We are in the process of testing the method on the lattice using non-interacting quarks.
It is interesting to compare our approach with the methods presented
in \cite{Blum:2015gfa}.  In the latter publication, the most
accurate results (for the fully connected contribution) were obtained
by effectively using position-space perturbation theory. There, rather
than $y$, it is the integral over the difference of the positions of two
quark-photon vertices, $r=x-y$, which needs to be performed. Due
to the use of a specific muon frame, that integral is
four-dimensional.  Also, the kernel is fully recomputed for every
value of $r$, which may cost a non-negligible fraction of the
computing time. While in \cite{Blum:2015gfa}, the kernel was computed on the same space-time 
lattice as the QCD four-point function, at this conference L.\ Jin presented a study where 
the kernel is computed on a larger volume, in order to reduce the power-law corrections in the volume.
There are thus a number of similarities, but also
significant differences between the methods of the two groups.
It is in any case encouraging that Blum et al.\ \cite{Blum:2015gfa} 
obtained a good signal for \(a_\mu^{\text{\hlbl}}\).
\medskip

We thank G.~von~Hippel and H.~Wittig for helpful discussions.
This work is supported in part by DFG through CRC 1044 ``The low-energy frontier of the Standard Model''.

\bibliographystyle{JHEP}
\bibliography{Nils_Asmussen-Lattice2016}

\end{document}